\documentclass[pra,twocolumn,showpacs,floatfix]{revtex4}
\usepackage{graphicx}
\usepackage[usenames]{color}
\newcommand{\beq}{\begin{equation}}
\newcommand{\eeq}{\end{equation}}
\newcommand{\beqa}{\begin{eqnarray}}
\newcommand{\eeqa}{\end{eqnarray}}
\newcommand{\ba}{\begin{array}}
\newcommand{\ea}{\end{array}}

\begin{document}

\title{Matter-wave localization in a random potential}

\author{Yongshan Cheng$^{1,2}$\footnote{yong\_shan@163.com}
and
   S. K. Adhikari$^1$\footnote{adhikari@ift.unesp.br;
URL: www.ift.unesp.br/users/adhikari}}
\address{$^1$Instituto de F\'{\i}sica Te\'orica, UNESP - Universidade
Estadual Paulista,
%Barra Funda,
01.140-070 S\~ao Paulo, S\~ao Paulo, Brazil\\
$^2$Department of Physics, Hubei Normal University, Huangshi 435002,
People's
   Republic of  China
}

\begin{abstract} By numerical and variational solution of the 
Gross-Pitaevskii equation, we studied the localization of a 
noninteracting and weakly-interacting Bose-Einstein condensate (BEC) in 
a disordered cold atom lattice and a speckle potential. In the case of a 
single BEC fragment, the variational analysis produced good results. For 
a weakly disordered potential, the localized BECs are found to have an 
exponential tail as in weak Anderson localization.  We also investigated 
the expansion of a noninteracting BEC in these potential. We find that 
the BEC will be locked in an appropriate localized state after an 
initial expansion and will execute breathing oscillation around a mean shape 
when a BEC at equilibrium in a harmonic trap is suddenly released into 
a disorder potential.

\end{abstract}


\maketitle

\section{Introduction}

Since the experimental realization of Anderson localization of a 
Bose-Einstein condensate (BEC) in a disorder potential 
\cite{Nature-453-891,Nature-453-895}, this topic has been the subject of 
intense theoretical and experimental activities. Billy {\it et al.} 
\cite{Nature-453-891} observed the exponential tail of the spatial 
density distribution when a $^{87}$Rb BEC was released into a 
one-dimensional (1D) waveguide in the presence of a controlled disorder 
created by a laser speckle. Roati \emph{et al}. \cite{Nature-453-895} 
employed a 1D quasi-periodic bichromatic optical lattice to 
observe  the Anderson localization of a noninteracting 
$^{39}$K BEC. Experimental studies employed bichromatic 
optical lattice \cite{PRL-98-130404,Nature-453-895}, shaken optical 
lattice \cite{PRA-79-013611}, cold atom lattice \cite{cal}
 and speckle potentials 
\cite{Nature-453-891}. (The effect of a repulsive atomic interaction on 
localization has also been studied \cite{experimental}.) 
Theoretical studies employed bichromatic optical lattice 
\cite{adhikari,PRA-80-053606,quasiperiodic2}, cold atom lattice 
\cite{PRL-95-020401,PRA-74-013616} and random 
\cite{PRA-79-063604,random2,NJP-10-045019} potentials
among others.

One can produce quasi-periodic \cite{quasiperiodic1} or random 
\cite{random1, PRL-95-070401, PRL-95-170409} potentials  for 
 Anderson localization by optical means. A 
bichromatic optical
lattice is realized by a primary lattice perturbed by a
weak secondary lattice with incommensurate wavelength \cite{Nature-453-895}.
Random speckle potentials 
are produced when
light is reflected by a rough surface or transmitted by a diffusive medium
\cite{chapter3}.
In addition 
to optical means, disordered or quasi-disordered potentials could also 
be created by using atomic mixtures \cite{PRL-95-020401} or 
inhomogeneous magnetic fields \cite{PRL-95-170401, JPB-39-1055}. Gavis and 
Castin 
\cite{PRL-95-020401} proposed a controlled way of 
producing a disordered potential 
by using a mixture of two different 
atomic species. 
%(Later, this idea
%has then been extended to three dimensions (3D)  \cite{PRA-74-013616}.)
In this approach, known as the cold atom lattice, 
the atoms of one of the two species, named  
``scatterers", are randomly trapped in the sites of an  optical 
lattice. If the filling factor is much less than unity, only some of the 
sites will be occupied by one atom and the others will be empty. The 
atoms of the second   species, 
that could be weakly affected by the presence of the 
optical lattice and denoted as ``test particles", feel the collisional 
interaction with the randomly-distributed scatterers. The collisional 
interaction will act as a random potential for the test particles.

Here, with variational and numerical solution of the Gross-Pitaevskii 
(GP) equation, we investigate the localization of cigar-shaped BECs in a 
1D random speckle potential and in a disordered cold atom lattice. The 
random speckle potential is modeled by taking identical Gaussian spikes 
distributed randomly \cite{NJP-10-045019}. In this case the spikes may 
overlap. The cold atom lattice is modeled by taking a periodic 
distribution of non-overlapping spikes from which a small number of 
spikes are taken out randomly \cite{PRL-95-020401}. We examine 
separately two regions of the localized BECs. The first region 
corresponds to the center, where the density profile of the localized 
BEC is quite similar to a Gaussian shape. In this region, we also use 
the variational approximation for some analytical understanding of the 
localized state. The second region corresponds to the tails of the BEC, 
where we focus on spatially extended wave functions with exponential 
decay corresponding to weak Anderson localization \cite{chapter3,NJP-10-045019}. 
The stability of the localized state is also investigated in both cases.
To the best of our knowledge this is the first detailed study of Anderson 
localization in the presence of only a disorder speckle or cold atom lattice 
potential. Most of the previous studies {\cite{NJP-10-045019,biol}}
employed  such a disorder potential 
superposed on a harmonic potential.

In Sec. \ref{II} we
present a brief account of the 1D GP equation, the disordered
potentials, and a variational analysis of the GP equation. In Sec.
\ref{III} we present numerical results using the split-step
Fourier spectral method and compare these with a variational
analysis. In Sec. \ref{IIII} we present a brief summary.

\section{Analytical consideration }

\label{II}

We assume that the trapping potential $V(x)$ is 
disordered along the longitudinal $x$ direction 
with  a strong harmonic trap  in transverse 
directions. Then it is appropriate to consider
a 1D reduction  of the three-dimensional
 GP equation by freezing the transverse
dynamics of the BEC to the respective ground states
and integrating over the transverse variables \cite{1d}. 
The BEC dynamics  of $N$ atoms can then be 
described by the following 1D equation  for wave 
function $u\equiv u(x,t)$   \cite{1d,1dd}
\begin{eqnarray}\label{eq1}
i\frac{\partial u}{\partial t}=- \frac{1}{2}\frac{\partial^2
u}{\partial x^2} +g|u|^2 u+V(x)u,
\end{eqnarray}
with nonlinearity $g= 2aN/a_\perp^2,$
normalization $\int_{-\infty}^{\infty}|u|^2 dx=1$, and  $a$
the atomic scattering length.   
The spatial
variable $x$, time $t$, and energy are expressed in transverse
harmonic oscillator units $a_\perp=\sqrt{\hbar/(m\omega)}$,
$\omega^{-1}$ and $\hbar\omega$, where $m$ is the mass of an atom
and $\omega$ is the angular frequency of the transverse trap.

The trap $V(x)$ can be modeled by a set of $S$ identical 
spikes randomly 
distributed along the $x$ axis \cite{NJP-10-045019}
\begin{eqnarray}\label{eq2}
V(x)=V_0\sum_{j=1}^S v(x-x_j),
\end{eqnarray}
where $V_0$ is the strength of the spike, and $v(x-x_j)$ represents
the potential of a single spike at position $x_j$. 
We consider Gaussian spikes \cite{NJP-10-045019}
\begin{eqnarray}\label{eq3}
v(x)=\left(\sigma \sqrt{\pi}\right)^{-1}\exp\left(-\frac{x^2}{\sigma
^2}\right),
\end{eqnarray}
with normalization $\int v(x)dx =1$ and width
$\sigma$. The statistical average of the disordered
potential (\ref{eq2}) is \cite{NJP-10-045019}
\begin{eqnarray}\label{eq4}
\langle V(x)\rangle\equiv \frac{1}{2L}\int_{-L}^L V(x) dx= \frac{V_0}{D},
\end{eqnarray}
with $D$ the average spacing between spikes and $2L$  the spatial 
extension of $V(x)$.  
The  auto-correlation function for potential $V(x)$ is defined as
\cite{NJP-10-045019}
\begin{eqnarray}\label{eq5}
C(d)=\langle V(x)V(x+d)\rangle-\langle V(x)\rangle^2,
\end{eqnarray}
where $\langle\cdot\rangle$ represents averaging as in Eq. (\ref{eq4}). 
An
average spike height $V_S$ is defined by  \cite{ NJP-10-045019}
\begin{eqnarray}\label{eq6}
V_S=\left[ \frac{1}{2L}\int_{-L}^{L}dx \left(V(x)-\langle
V(x)\rangle\right)^2\right] ^{1/2}.
\end{eqnarray}

In presence of  strong disorder  a
multi-fragmented BEC state is expected 
 \cite{PRA-79-063604}. Under appropriate conditions,
however, the shape of localized states may be a single fragment, 
so the
variational treatment will be useful   \cite{CYS}. 
The variational approach  
allows to obtain useful relations among the localized state parameters. 
We consider the stationary form $\phi(x)$ of the localized state 
given by $u(x, t)=\exp(-i\mu t)\phi(x)$, with $\mu$ the chemical
potential. The real wave function, $\phi(x)$, obeys 
\begin{equation}\label{eq7}
\mu\phi(x)+\phi''(x)/2-g\phi^3(x)-V(x)\phi(x)=0,
\end{equation}
where the prime denotes space derivative. 
We use the
off-center variational Gaussian ansatz
\begin{eqnarray}\label{eq8}
\phi(x)=\frac{1}{\pi^{1/4}}\sqrt{\frac{{\cal N}}{w}}
\exp\left[-\frac{\left(x-x_0\right)^2} {2w^2}\right] ,
\end{eqnarray}
with $w$ the width, 
 $x_0$  the center and ${\cal N}$  the normalization
of the localized BEC. The
Lagrangian of the system is 
\begin{eqnarray}\label{eq9}
{\cal L}&=&\int_{-\infty}^{\infty}\left[\mu\phi^2-(\phi ')^2/2-g\phi^4/2
-V(x)\phi^2  \right]dx -\mu ,  \nonumber  \\
 &=& \mu ({\cal N}-1)-\frac{{\cal N}}{4w^2}- \frac{g{\cal
N}^2}{2\sqrt{2\pi}w}-\sum_{j=1}^S
 {\cal N}{\cal L}_{j},\\
{\cal L}_j&\equiv& \frac{
V_0}{\sqrt{\pi\left(\sigma^2+w^2\right)}} 
\exp\left[-\frac{\rho_j^2}{\sigma^2+w^2}\right],
\end{eqnarray}
where  $\rho _j=x_0-x_j$. 
The Euler-Lagrange equation  $\partial {\cal L}/\partial \mu =0$ yields
the normalization ${\cal N}=1$. We  use it in the following
equations. The remaining equations $\partial {\cal L}/\partial w = \partial
{\cal L}/\partial {\cal N}=0$ yield, respectively,
\begin{eqnarray}\label{eq10}
1&=&-\frac{gw}{\sqrt{2\pi}}+\frac{2w^4}{\left(\sigma^2+w^2\right)}
 \sum_{j=1}^S\left(\frac{2\rho_j^2}{\sigma^2+w^2}-1\right){\cal L}_j,\\
\mu&=&\frac{1}{4w^2}+\frac{g}{w\sqrt{2\pi}}+\sum_{j=1}^S {\cal L}_j,
\label{eq11}
\end{eqnarray}
and determine the width $w$ and the chemical potential $\mu$.
The corresponding energy $E=
\int_{-\infty}^{\infty}[(\phi')^2/2+g\phi^4/2+V(x)\phi^2]dx$ is
given by
 \begin{eqnarray}\label{eq12}
E=\frac{1}{4w^2}+\frac{g}{2w\sqrt{2\pi}}+\sum_{j=1}^S {\cal L}_j.
\end{eqnarray}

\section{Numerical Results}

\label{III}

We perform the numerical integration of  GP equation (\ref{eq1}) 
employing the imaginary- or real-time split-step Fourier spectral method 
with space step 0.04, time step 0.001 and an  initial input Gaussian pulse.
The time evolution is continued till convergence.  We checked the accuracy of 
the results by varying the space and time steps and the total 
number of space and time steps. Although we use a time-dependent 
approach, the localized states are stationary.

\subsection{A speckle potential}

\begin{figure}%[!ht]
\begin{center}
\includegraphics[width=.49\linewidth]{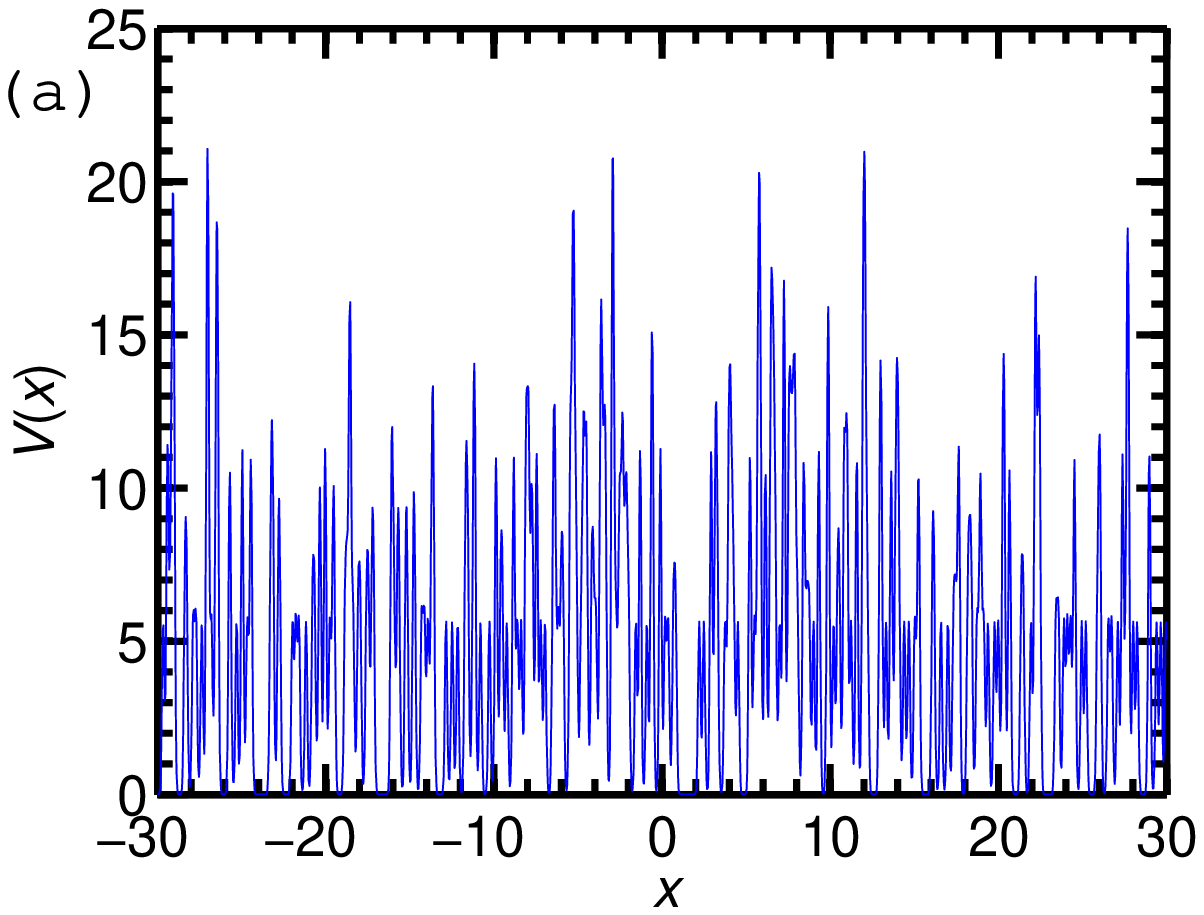}
\includegraphics[width=.49\linewidth]{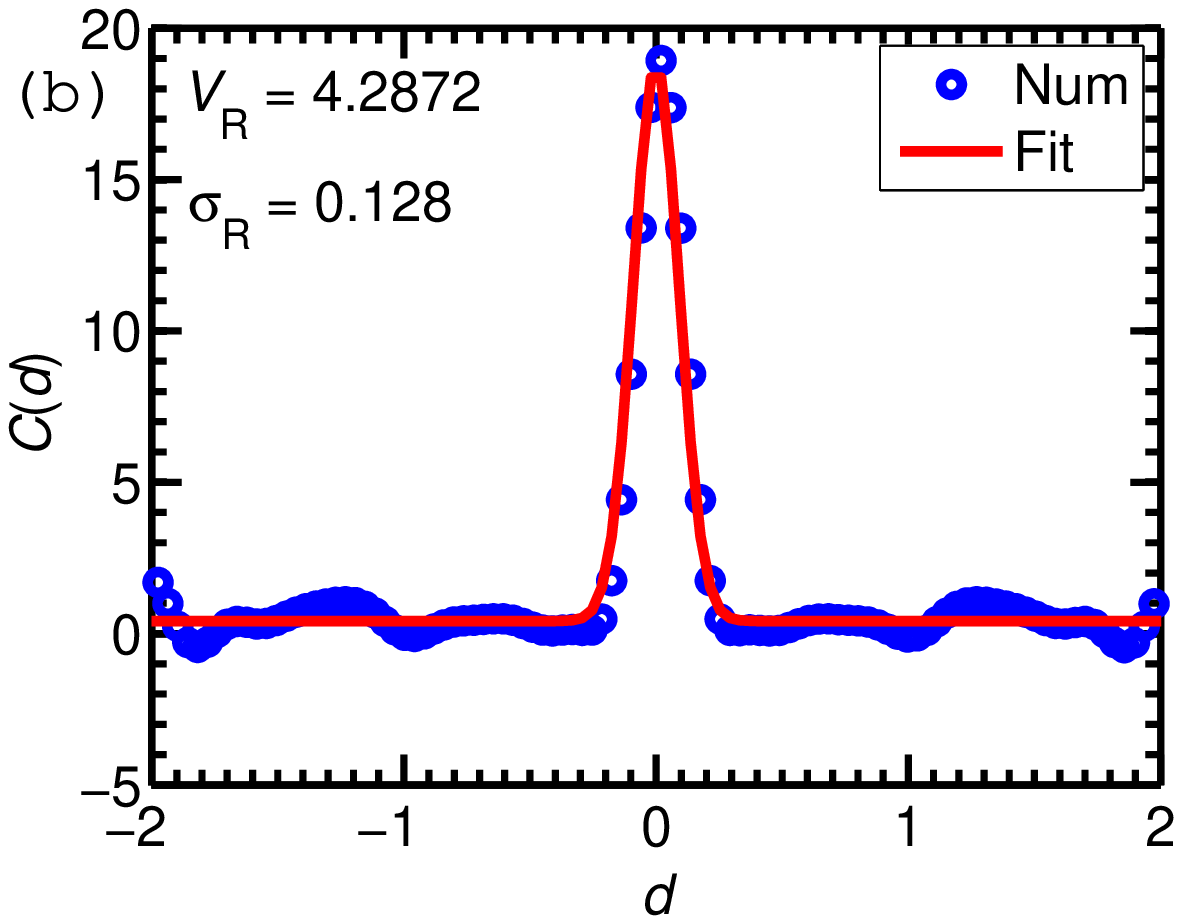}
\end{center}

\caption{(Color online) (a) Typical disordered speckle potential
from Eqs. (\ref{eq2}) and (\ref{eq3}) with  
$S=300,
L=30, V_0=1$ and $\sigma=0.1$. (b) The  auto-correlation
function (\ref{eq5})
of the potential (a). The circles (Num)
denote numerical results from Eq. (\ref{eq5}) and the
solid line (Fit) is the  Gaussian fit  (\ref{13}) with $V_R=4.2872$
and $\sigma_R=0.128$
 } \label{fig1}
\end{figure}

To generate the speckle potential 
we use a set of
random numbers with the standard MATLAB function RAND \cite{NJP-10-045019}. 
These numbers
are then mapped
into the interval $[-L, L]$ by a linear transformation and 
denote the position $(x_j)$ of the spikes. We select
$S=300, L=30$ and a small width
$\sigma=0.1$.
The spikes are described by Eq. (\ref{eq3}) and the speckle potential by 
Eq. (\ref{eq2}). 
A typical speckle potential 
{$V(x)$} for $V_0 = 1$
is plotted in Fig.   \ref{fig1} (a). 
The statistical average of Eq. (\ref{eq4}) is 
$\langle V\rangle=5$. 
 Using Eq. (\ref{eq6})
the average speckle
height is $V_S=4.3519$, 
and the spatial auto-correlation function (\ref{eq5})
is
shown in Fig. \ref{fig1} (b) where we also show the numerical fit 
\cite{NJP-10-045019}
\begin{equation}\label{13}
C(d)\approx V_R^2\exp(-d^2/\sigma_R^2),
\end{equation}
with amplitude $V_R=4.2872$  and correlation
length $\sigma_R=0.128$.
We study localization with 
potential $V(x)$ of Fig. \ref{fig1} (a) 
for  different $V_0$. 
This potential can  produce a stationary BEC 
 near  $x=0$. 
We also repeated our studies  with different
speckle potentials generated with the same method and parameters as
$V(x)$ but by different random processes. Similar conclusions are 
obtained in all cases 
except that the center of the localized states could  be 
different.

\begin{figure}%[!ht]
\begin{center}
\includegraphics[width=0.7\linewidth]{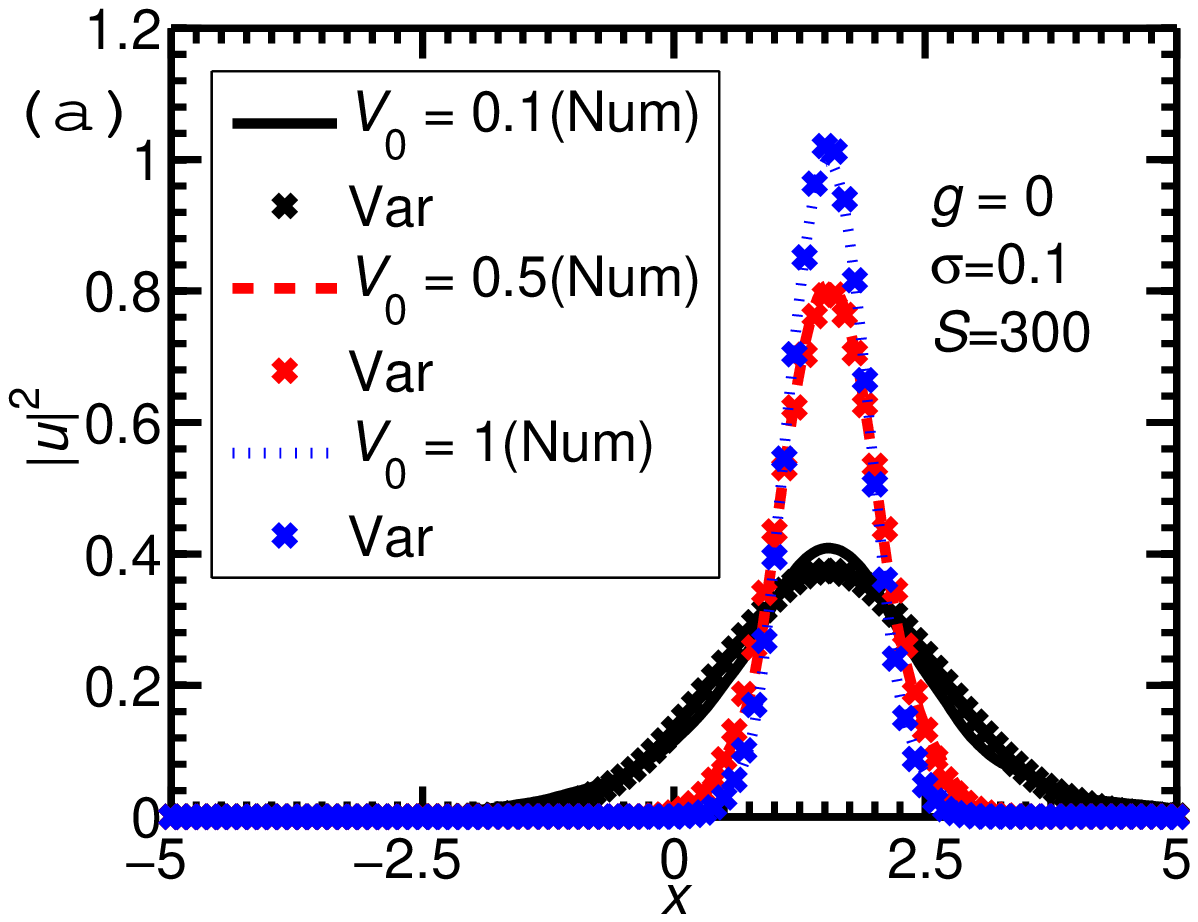}\\
\includegraphics[width=0.7\linewidth]{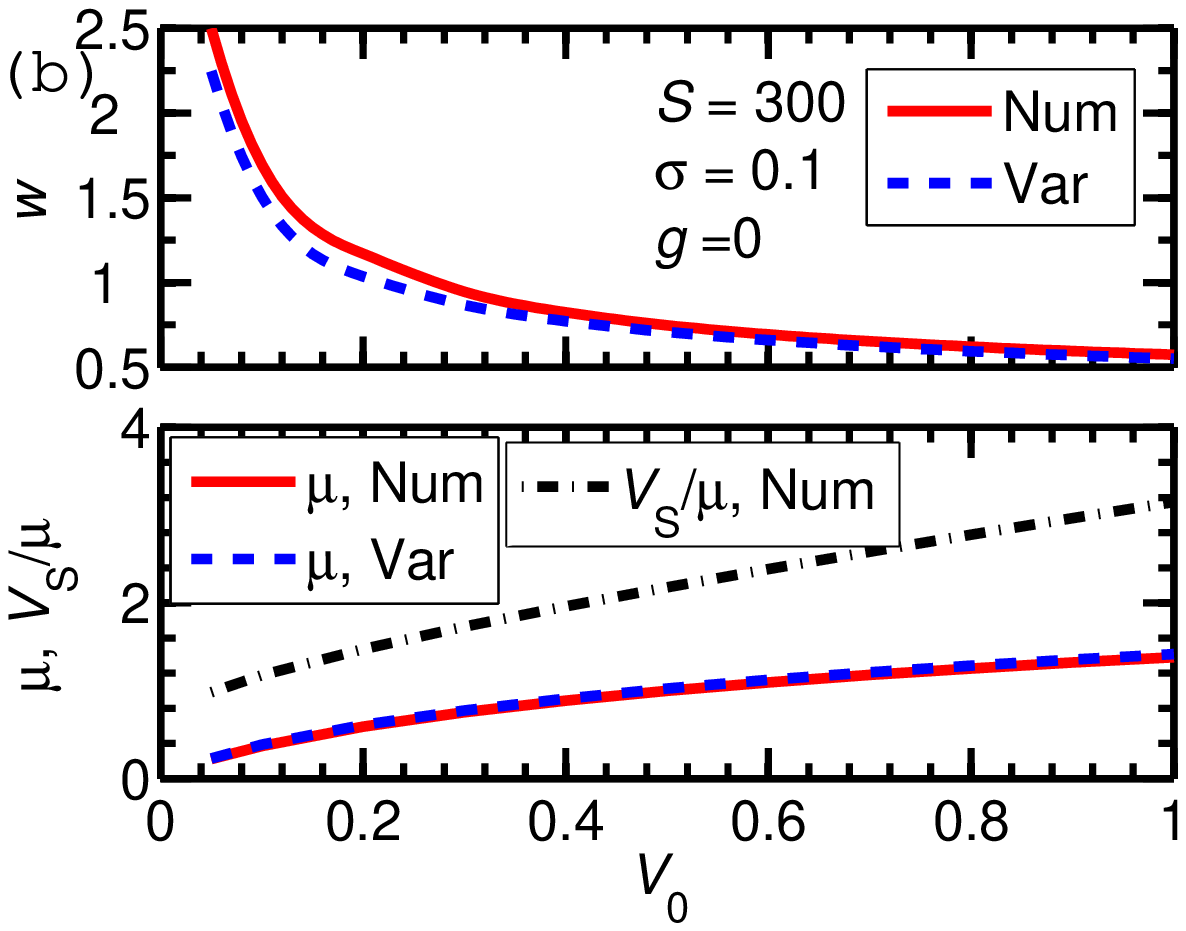}\\
\end{center}

\caption{(Color online)  (a) The numerical (Num)
and variational (Var) densities $|u|^2$
of the localized
BEC  versus $x$ for different $V_0$. 
 (b) The numerical and  variational  
widths $w$, chemical potential $\mu$ and $V_S/\mu$  
of the  BEC versus $V_0$. } \label{fig2}
\end{figure}

We first investigate the central region 
of the localized 
BEC, and the numerical and variational results are exhibited  in
Figs.  \ref{fig2} (a) and (b) where 
the BEC center $x_0$ is obtained by a Gaussian fitting of the
numerical 
density profile.  In Fig.  \ref{fig2} 
(a) we present the
density profiles  $|u|^2$
of the localized BEC  for
different $V_0$. Good agreement between variational and numerical 
densities is obtained. 
To
understand the effect of $V_0$, the numerical  
and variational 
widths (upper pannel)
and the chemical potential $\mu$ and $V_S/\mu$ (lower pannel)
of
the localized BEC versus $V_0$ are plotted in Fig.   \ref{fig2} (b) for $g=0$.
(The numerical width is the root mean square size of the BEC.)
Figure \ref{fig2}
(b) shows that the width decreases and the chemical potential
increases with the increase of $V_0$. However, the rate of 
change is smaller when
$V_0$ is large enough (for example, $V_0 >0.4$). It means that the
kinetic energy dominates in the regime. The quantity $V_S/\mu$ gives a
measure of disorder. However, the values of $V_S/\mu$ cannot 
be directly compared with those in Ref. \cite{NJP-10-045019} 
where a harmonic potential 
is used together with the disorder potential, so that a smaller disorder 
may lead to localization because of the additional harmonic potential.

\begin{figure}%[!ht]
\begin{center}
\includegraphics[width=.49\linewidth]{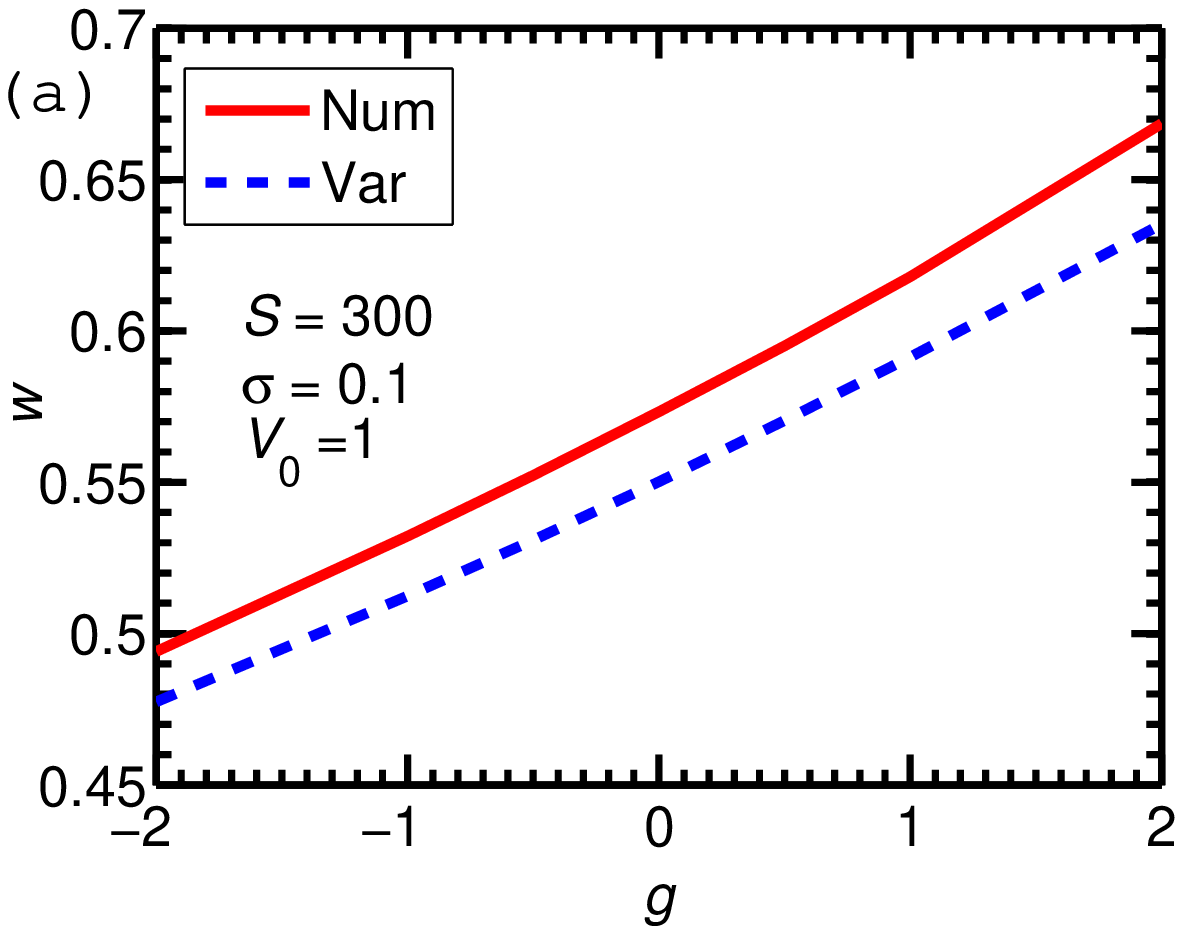}
\includegraphics[width=.49\linewidth]{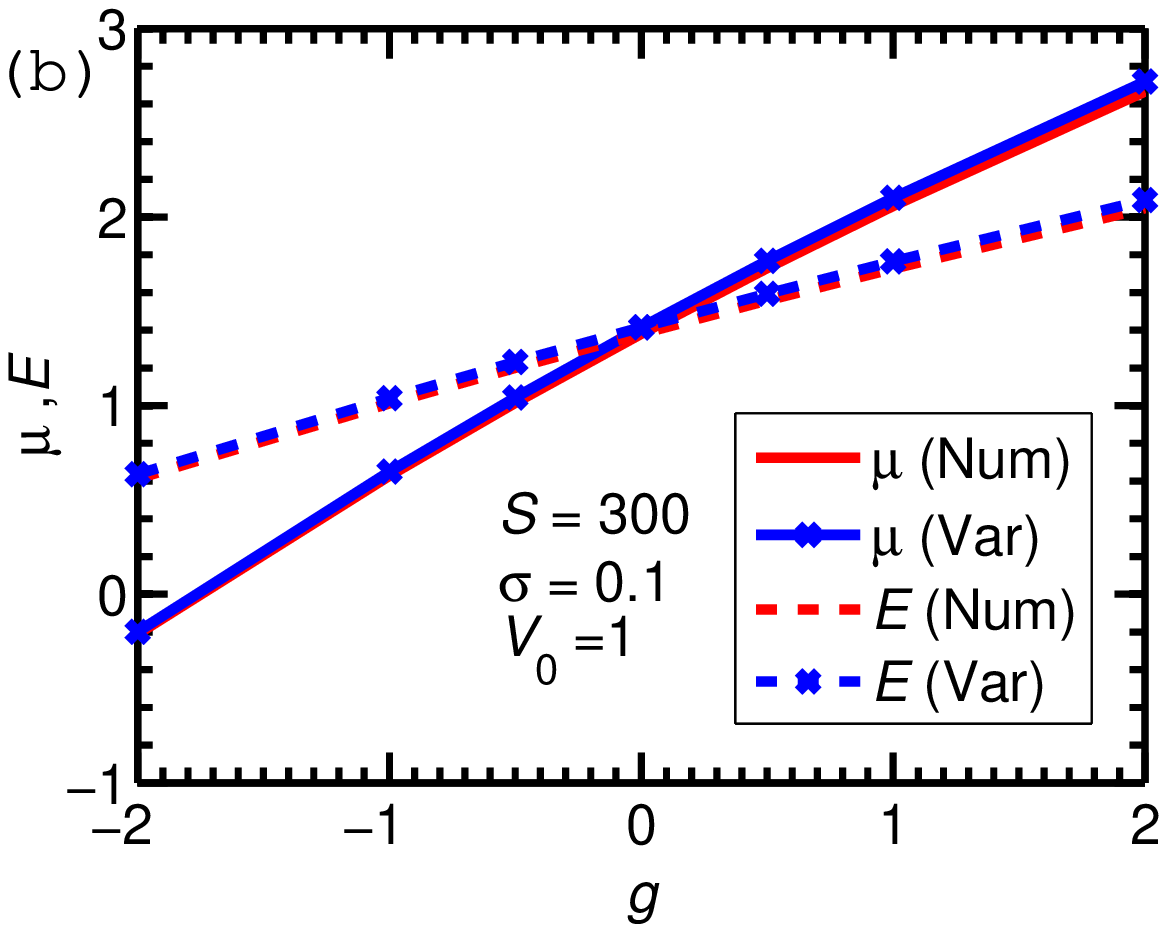}
\end{center}

\caption{(Color online)  The effects of $g$ on the center region of
the BEC. (a) The width and  (b) the energy $E$ and chemical
potential $\mu$ versus  $g$. }
\label{fig3}
\end{figure}

Next we consider the effects of a small nonlinearity $g$ on the
localized BEC in the case of a stronger disorder (viz. $V_0=1$). The
width, energy  and chemical potential of the localized BEC versus $g$ are
plotted in Figs.   \ref{fig3} (a) and (b).
The width,  energy  and chemical potential
 increase with   increasing nonlinearity.
Increasing nonlinearity means increasing repulsion and hence 
increased width. Increasing nonlinearity also means a larger 
positive contribution to 
the Hamiltonian and hence a larger  energy  and chemical potential.

\begin{figure}%[!ht]
\begin{center}
\includegraphics[width=.7\linewidth]{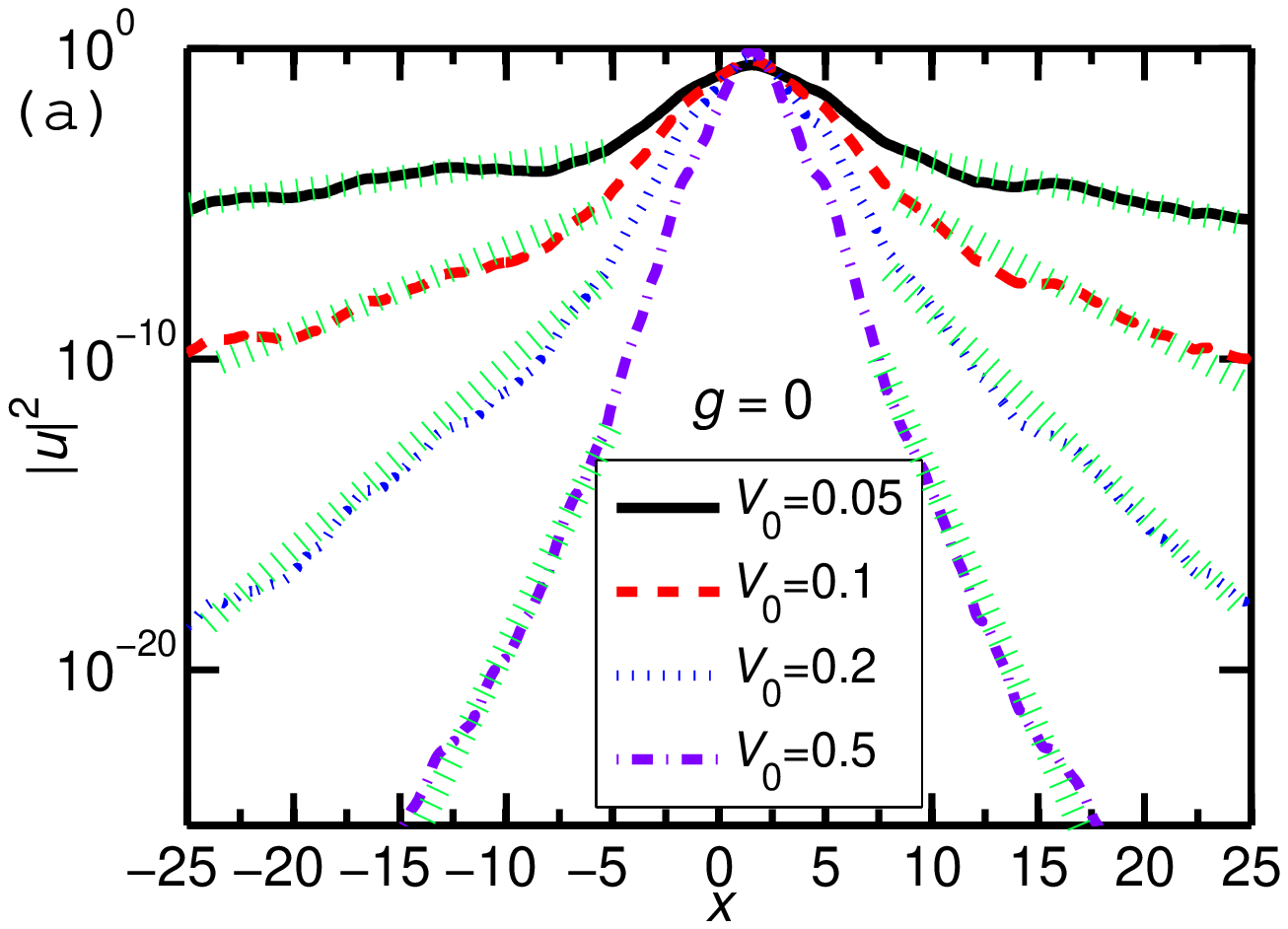}\\
\includegraphics[width=.7\linewidth]{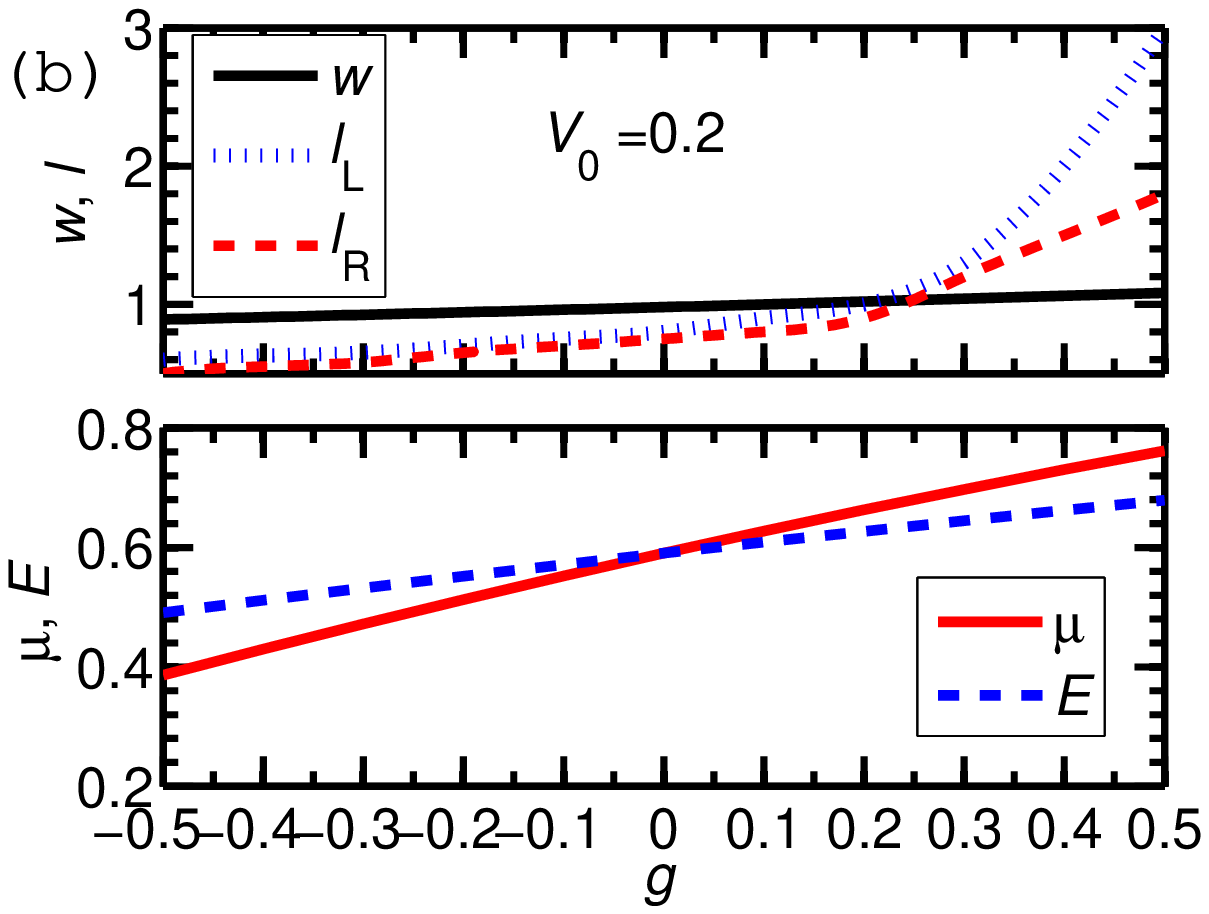}\\
\end{center}

\caption{(Color online)  (a) The numerical density $|u|^2$
and its exponential fit 
(hatched line) versus $x$ for different $V_0$ and $g=0$.
(b) (the upper panel) The numerical width $w$ 
 and the localiztion lengths $l_L$ (left tail) and $l_R$ (right tail)
versus the nonlinearity $g$ for $V_0=0.2$. (the lower panel) The numerical 
chemical potential $\mu$ and energy $E$ versus  $g$ for $V_0=0.2$.} 
\label{fig4}
\end{figure}

So far, we investigated the central region of the localized BEC. In the 
presence of strong disorder, we also studied the tail region of the BEC 
but did not find an exponential decay. However, one of the more 
interesting issues of localization is in the presence of a weak disorder 
when the system is localized due to the disorder nature of the potential 
but not due to the strength of the potential. In the presence of a weak 
disorder, Anderson-like localization is expected in the tail region of 
the BEC. The weak disorder of a speckle potential can be acquired in two 
ways. One way is to reduce the strength $V_0$ in the potential $V(x)$. 
Another way is to reduce the disorder of the spikes' position. (This 
second case will be discussed in the following subsection). Firstly, we study 
numerically the tail region of the BEC in potential $V(x)$ with smaller 
$V_0$. Figure \ref{fig4} (a) shows the effects of $V_0$ on the tail 
region of the BEC for $g=0$. The hatched lines are exponential fitting to 
the tails by $ \exp(-|x-x'|/l)$. Where $l$ is the localization 
length \cite{Nature-453-891,NJP-10-045019}
and $x'$ is the center  of the exponential function. The 
density profile obtained numerically is found to have a clean exponential 
tail for smaller $V_0$. When $V_0$ is large enough, the exponential 
tail fades away and the tail region approximates the central Gaussian 
distribution. We also find that the tail is modulated by the speckle 
potential if $V_0$ is small enough. In general, the center of the 
exponential function is different from the center of the localized 
state. Also, the density profile of the tail is asymmetric around the 
central position because of the asymmetry of the disordered potential. 
Thus, the localization length for the left tail $(l_L)$ is different 
from that for right tail $(l_R)$. The stronger the disorder is, the 
smaller the localization length is. For $V_0<0.2$ or for $V_S < 
0.86$ corresponding to $V_S/\mu < 1.5$ (viz. Fig. \ref{fig2} (b)), 
we have weak Anderson 
localization for $g=0$. These values of $V_S/\mu$ are larger than $V_S/\mu 
<0.3$ considered in Ref. \cite{NJP-10-045019}. This is reasonable, as
in Ref. \cite{NJP-10-045019} the authors consider a speckle 
potential superposed on a harmonic potential.

We study numerically the effect of the small nonlinearity $g$ on the
localization of  BEC as well. The results are presented in Fig. 
\ref{fig4} (b), where  
the numerical 
width $w$ (corresponding to the center region) and the localiztion 
lengths $l_L$ (left tail) and $l_R$ (right tail)  versus nonlinearity $g$ are 
shown in the upper panel. When
$g$ is smaller, the two localization lengths 
are  similar as shown
in the upper panel in Fig.  \ref{fig4} (b). If $g$ is larger, the asymmetry
between the two tails becomes larger. The localization 
lengths increase sharply
with $g$ for  $g>0.3$. The localization is  destroyed
and the BEC becomes  a multi-fragmented state 
if $g >0.5$ and $V_0 =0.2$.
Compared with Fig. \ref{fig3} (b), the lower panel in Fig. 
\ref{fig4} (b) shows that the
energy is smaller when $V_0$ is smaller. The result is accordance
with Fig.  \ref{fig2} (b).

\begin{figure}%[!ht]
\begin{center}
\includegraphics[width=.49\linewidth]{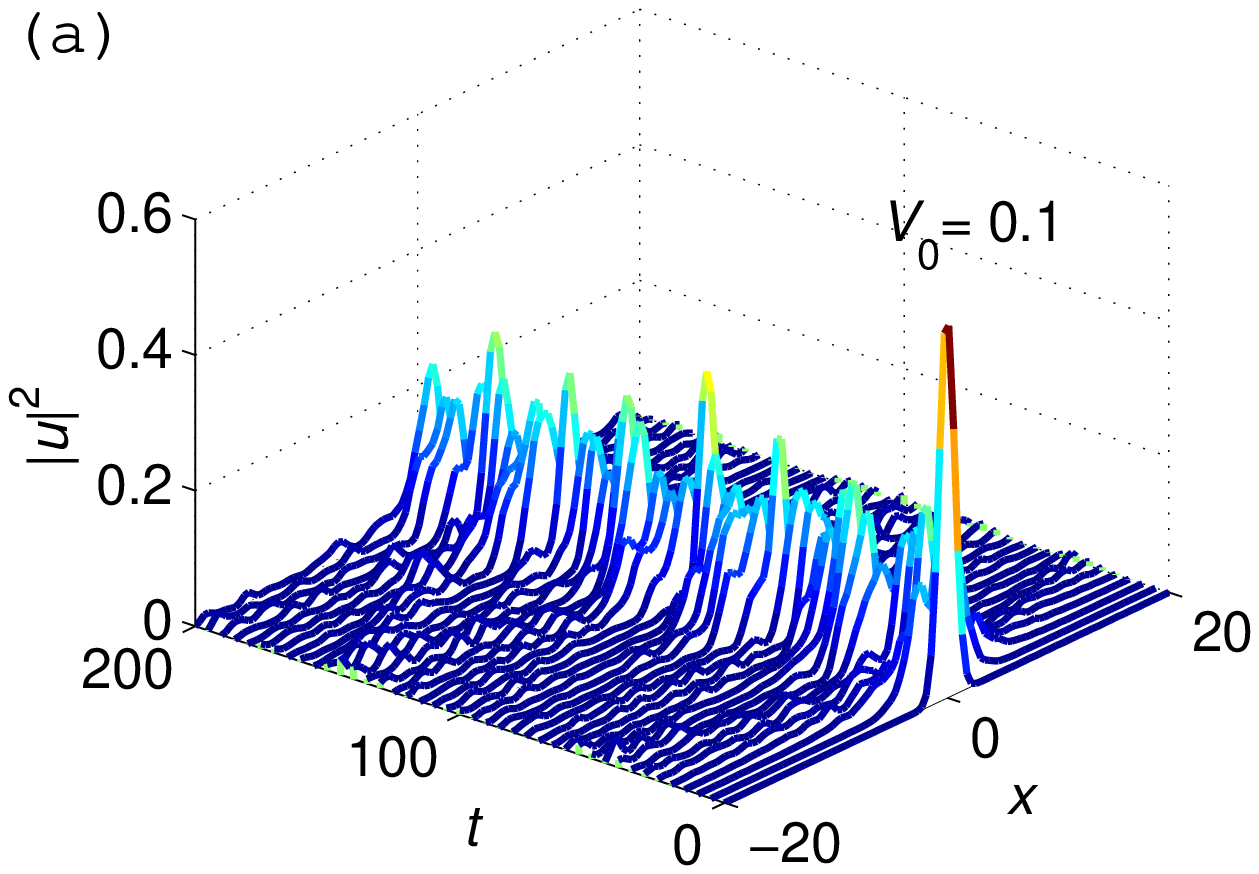}
\includegraphics[width=.49\linewidth]{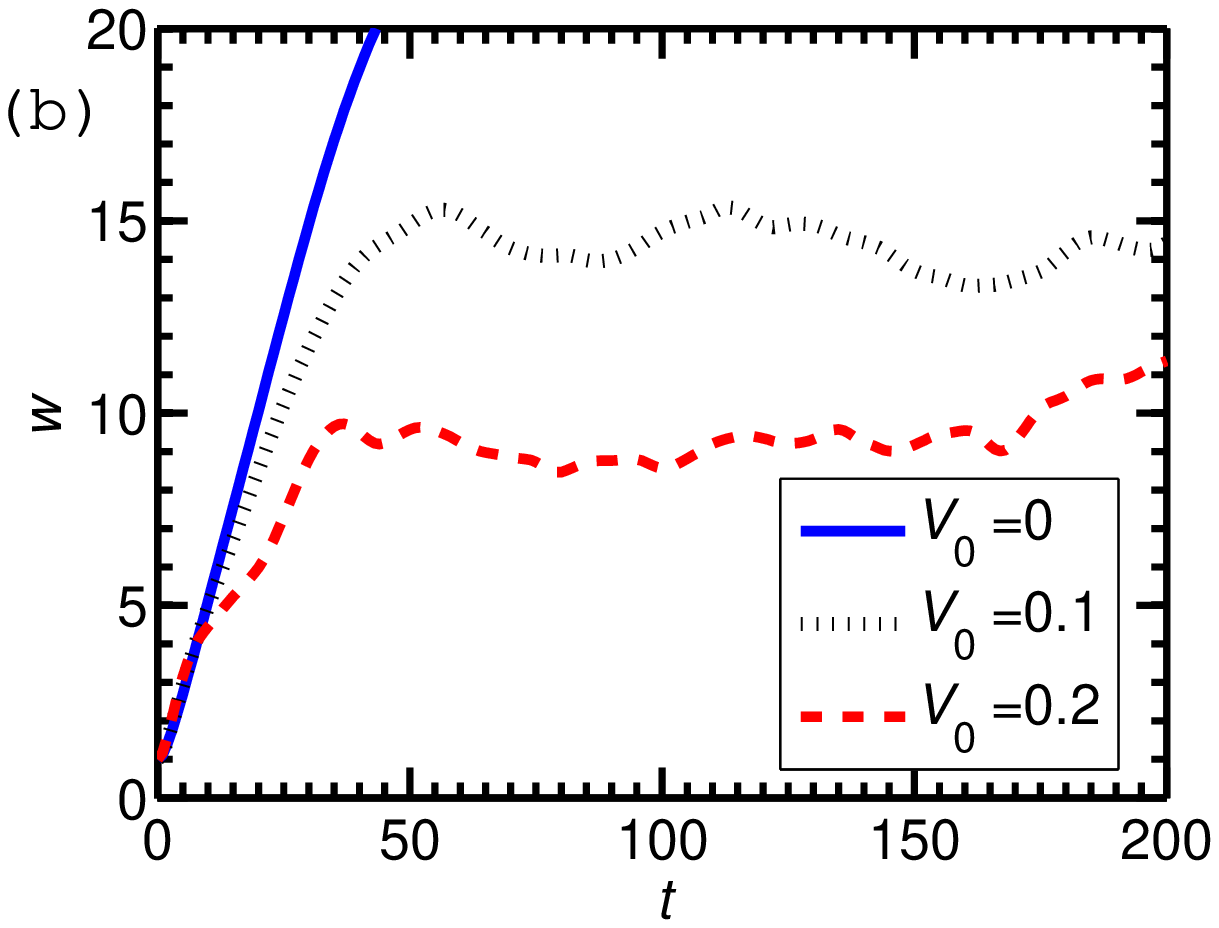}
\end{center}

\caption{(Color online) The expansion of a noninteracting BEC in the
speckle potential. (a)  The time evolution for the density profile 
$|u|^2$ versus $x$ and $t$
of the
BEC for $V_0=0.1$. (b) The time evolution of the width of the
BEC for several $V_0$. } \label{fig5}
\end{figure}

We also studied the expansion of a noninteracting BEC in the
speckle potential. We start from
a BEC at equilibrium in the harmonic trap. At time $t = 0$, the
harmonic trap is suddenly changed into a weak speckle potential as
 in the experiment of Billy \emph{et al}.
\cite{Nature-453-891}. The evolution for the density profile of the
BEC is shown in Fig. \ref{fig5} (a) 
where the initial expansion and the
consequent breathing oscillation of the BEC are illustrated. During
the evolution, the BEC remains localized  and oscillates
around an equilibrium point. To understand this behavior, the
time evolution of the width of the BEC is plotted in 
Fig. \ref{fig5} (b)
for several $V_0$ corresponding to a weak disorder. During the
initial expansion, the BEC expands quickly. If $V_0=0$, the BEC
cannot be localized and expands into the whole space (see the
solid line). If the potential has weak disorder, however, it will be
locked in an appropriate localized state after a certain amount of
expansion. After this happens, the system executes breathing
oscillation around a mean shape of the localized state and the 
width remains bounded and does not increase indefinitely with time.
The stronger $V_0$ is, the shorter the time for the initial
expansion is and smaller is the final width.

\begin{figure}%[!ht]
\begin{center}
\includegraphics[width=.7\linewidth]{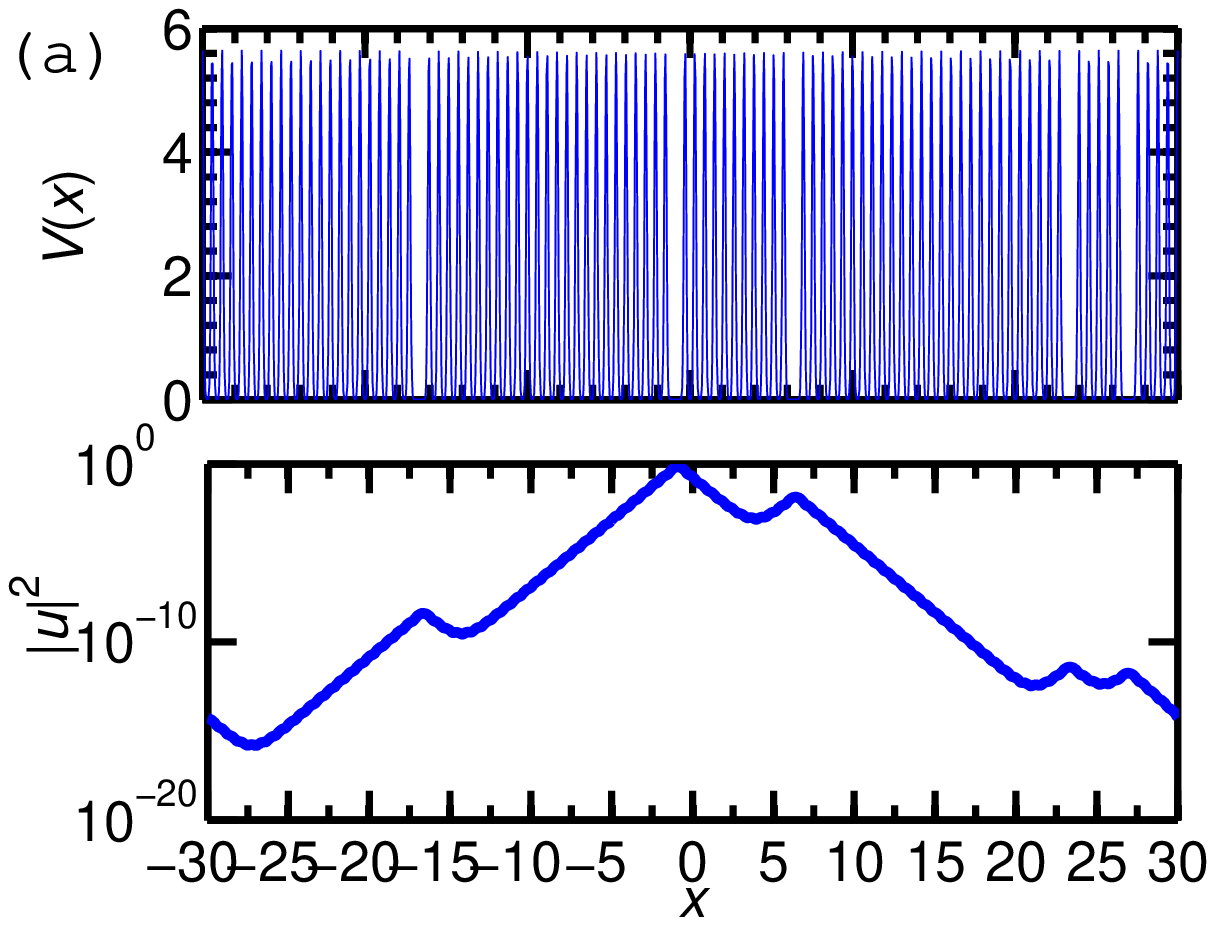}\\
\includegraphics[width=.7\linewidth]{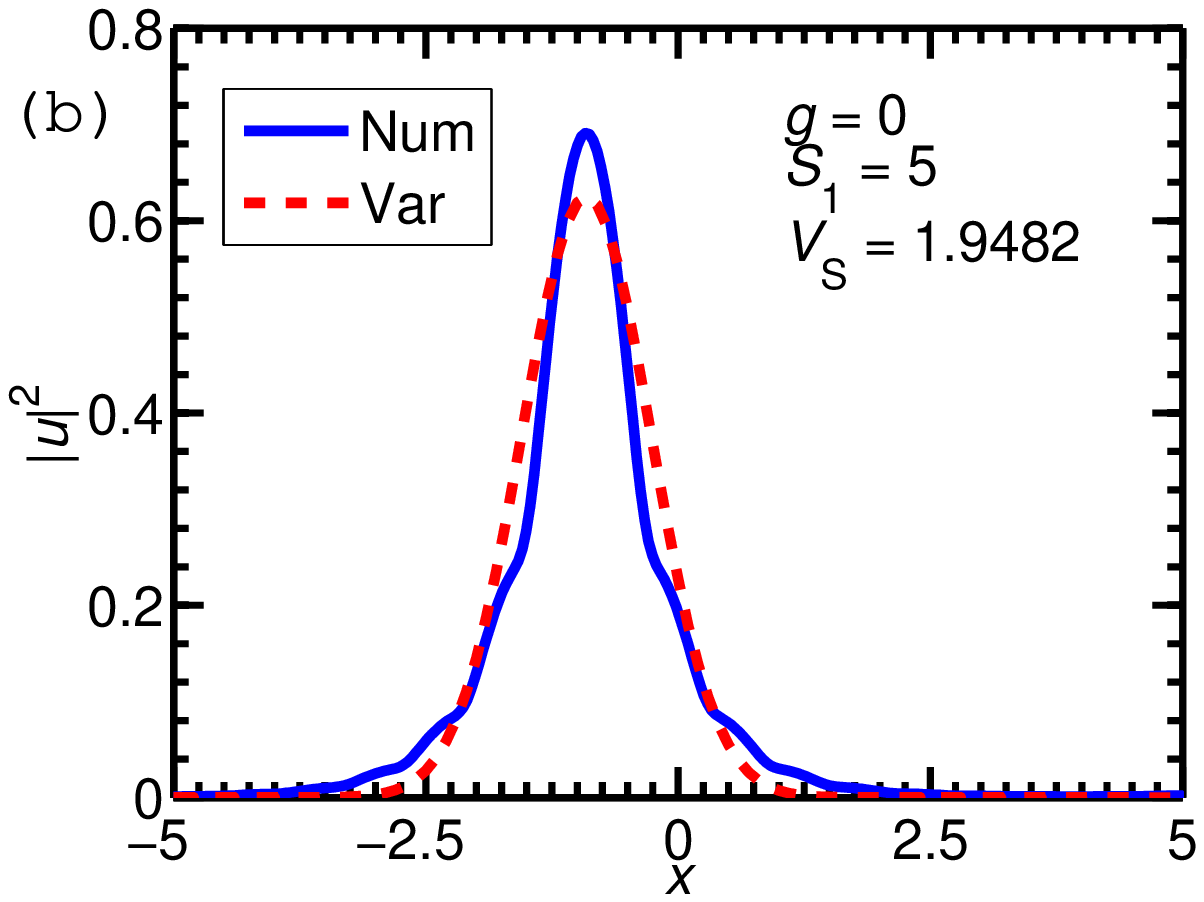}\\
\end{center}

\caption{(Color online) (a) A typical   cold atom
lattice (upper panel), and the numerical density profile
of the localized state (lower panel) in exponential scale. 
(b) The numerical (Num) and variational (Var) density
profiles $|u|^2$ of the localized BEC with 
disordered potential  (a). } \label{fig6}
\end{figure}

\subsection{A disordered cold atom lattice}

Following the model given by Gavish and
Castin \cite{PRL-95-020401}, the disordered cold atom lattice is taken as 
a set of
periodical spikes with a few  random vacancies (holes). To numerically 
generate such a potential, we
consider  $S=100$ periodic  spikes within $x=[-30,
30]$ defined by Eq. (\ref{eq3}) with $\sigma=0.1$
and $V_0 = 1$. Then we take out a small number $S_1$ of 
spikes whose positions are
obtained with the standard MATLAB function RAND. Figure  \ref{fig6} 
(a) shows a
typical weakly disordered potential with $S_1=5$ (upper panel), and
the numerical density profile of the localized BEC
in  this potential (lower panel) for $g = 0$.
The modulation of the 
exponential tail of the density profile is clearly visible. 
In  Fig. \ref{fig6} (b) we compare the central part of the numerical 
density  with the  
 variational density for 
the potential of  Fig. \ref{fig6}  (a).

\begin{figure}%[!ht]
\begin{center}
\includegraphics[width=.49\linewidth]{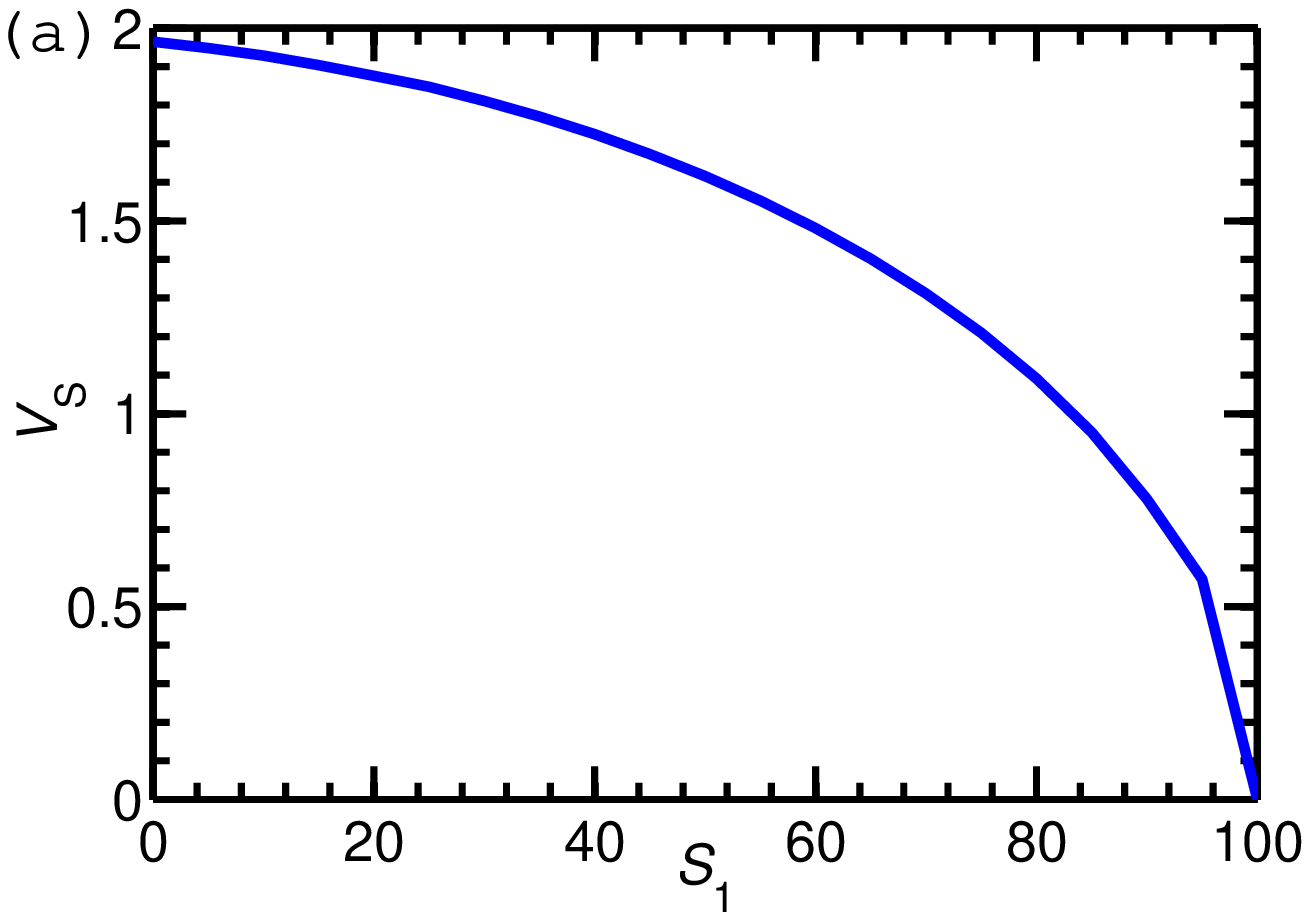}
\includegraphics[width=.49\linewidth]{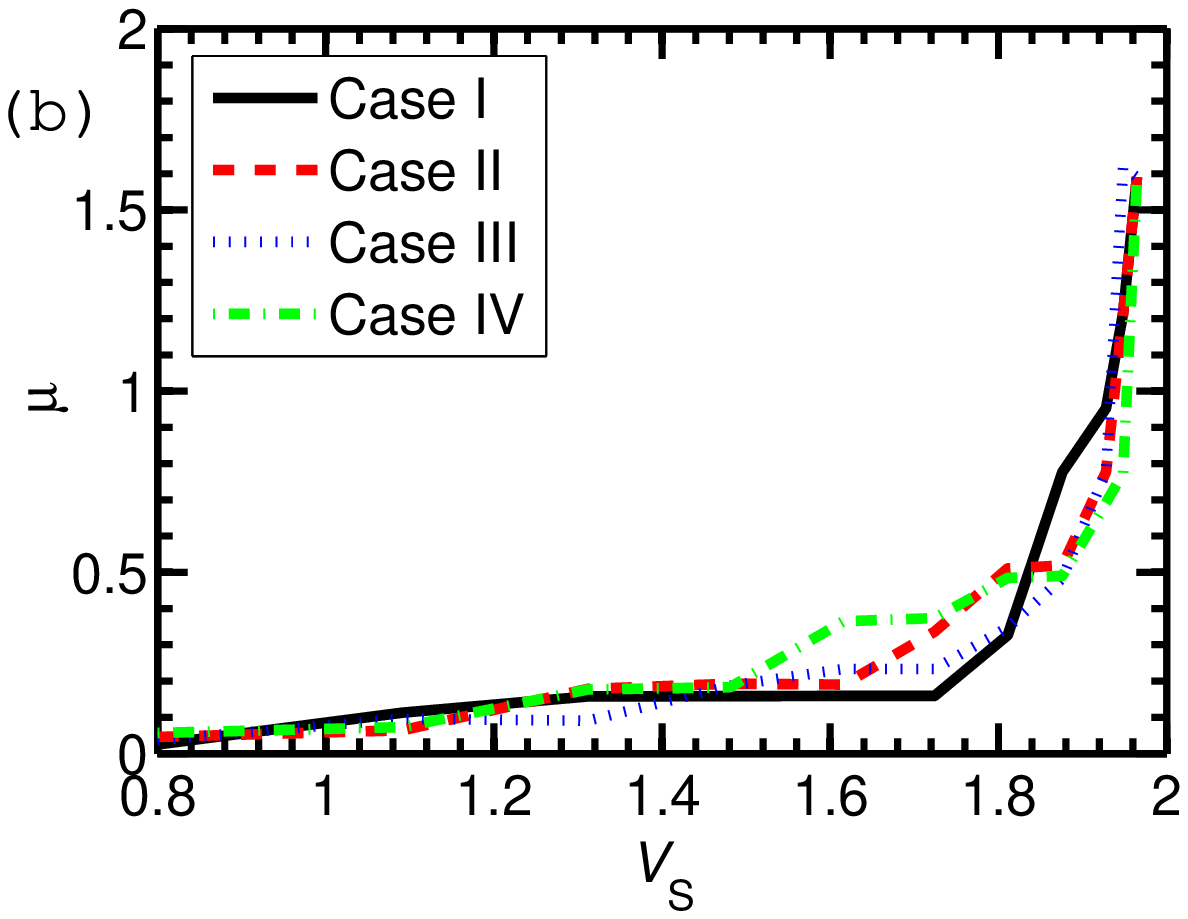}
\end{center}

\caption{(Color online) (a) The average spike height  of the disordered
potential $V_S$ of Eq. (\ref{eq6}) versus $S_1$. (b) The chemical potential of the
stationary localized states versus  $V_S$ in case of four cold atom lattice. }
\label{fig7}
\end{figure}

In Fig. \ref{fig7} (a) we plot the average spike height $V_S$ of Eq. 
(\ref{eq6}) versus $S_1$ showing a decrease of $V_S$ with $S_1$.
To understand the effects of the
average speckle height on the  localized BEC, we numerically
calculate the chemical potential of the localized state
versus $V_S$ of Eq. (\ref{eq6}) 
for four cases. The disordered cold atom lattices
corresponding in these cases  are created by the same parameters but with
different random processes. Although the disordered potentials
created by the different random processes possess different local
characteristics, the global effects are quite similar. The value and
variation  of the chemical potential of the system are smaller if $V_S$
is smaller, corresponding to strong disorder.
 In the case of the weak disorder, for example $V_S
>1.8$, the chemical potentials are very sensitive to the average
speckle height. In this case a large $V_S$ corresponds to weak disorder 
in contrast to the speckle potential of Fig. \ref{fig1}, where the same denotes 
strong disorder.

\begin{figure}%[!ht]
\begin{center}
\includegraphics[width=.49\linewidth]{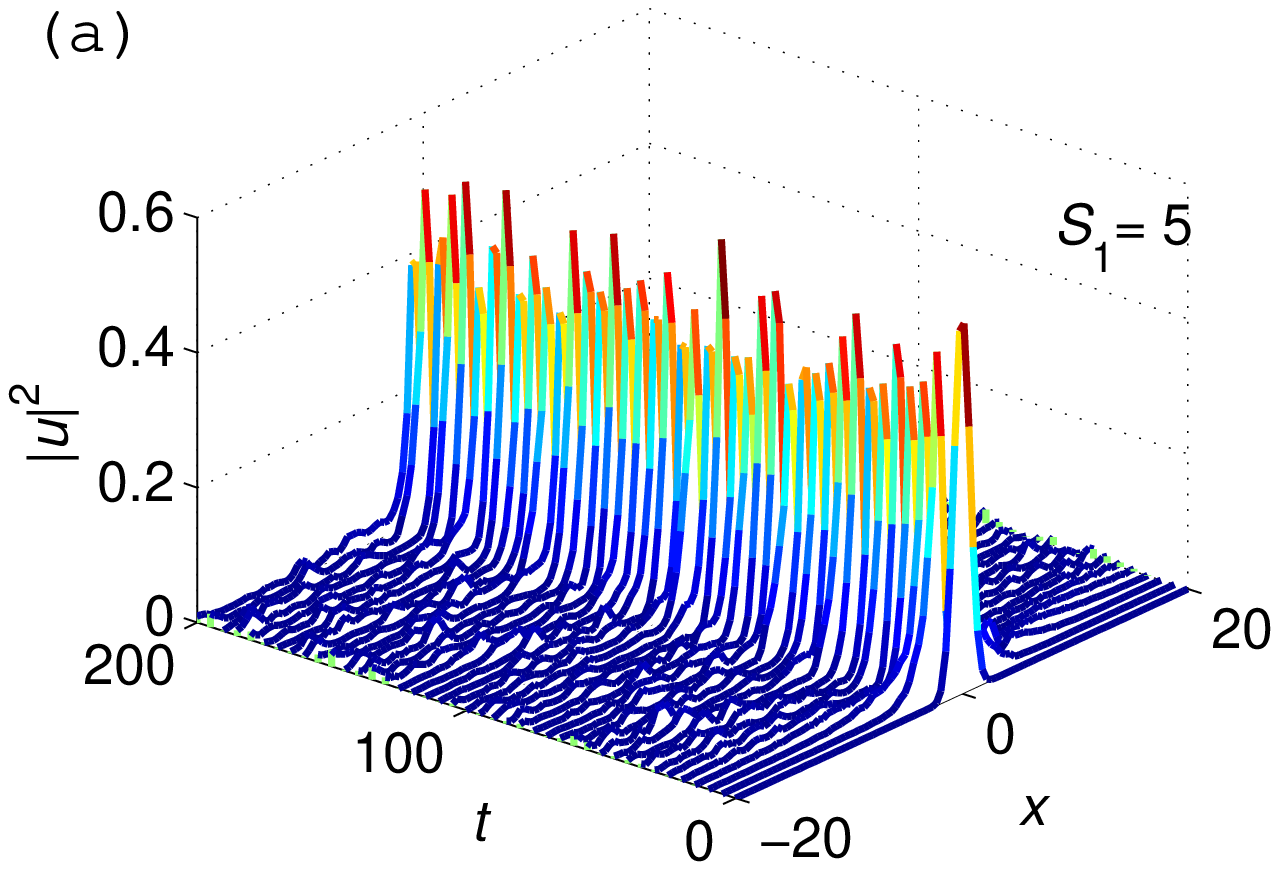}
\includegraphics[width=.49\linewidth]{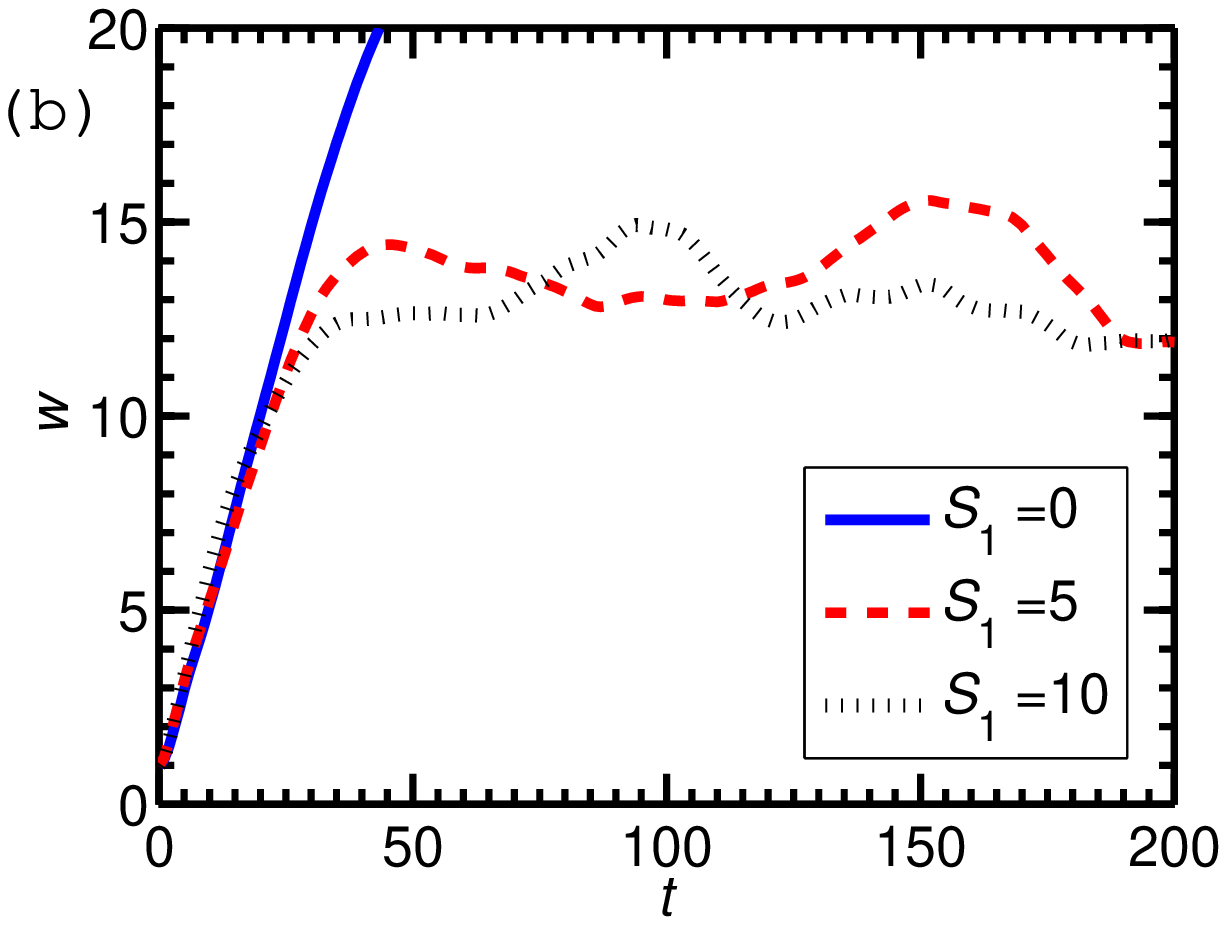}
\end{center}

\caption{(Color online) The expansion of a noninteracting BEC in the
weak disordered cold atom lattice. (a) The evolution of the
density profile of the BEC for $S_1=5$. (b) The  evolution of
the width of the BEC for several $S_1$.
 } \label{fig8}
\end{figure}

We also study the dynamics of the localized BEC trapped in a
weak disordered cold atom lattice. An harmonically trapped BEC is 
suddenly released into a weak disordered cold atom lattice shown by Fig. 
\ref{fig6} (a).  The evolution of the density profile and width of 
the BEC are plotted in Figs.  \ref{fig8} (a) and (b), respectively. 
Because of the weak disorder, the BEC can remain as a single fragment 
during the evolution (see Fig.  \ref{fig8} (a)). If the potential is 
periodical (corresponding to $S_1=0$), the BEC can not be localized. For 
a small $S_1$ the width executes stable breathing oscillation 
indicating stable localization.

\section{Summary}
\label{IIII}

In this paper, using the numerical and variational solution of the 
time-dependent GP equation, we studied the localization of a 
noninteracting and weakly interacting BEC in a disordered potential. We 
considered two models of the disordered potential corresponding to (i) 
the speckle potential \cite{NJP-10-045019,Nature-453-891} and (ii) the 
disordered cold atom lattice \cite{NJP-10-045019,PRL-95-020401}.  These 
disordered potentials are created by a superposition of narrow spikes 
randomly distributed in space. In the case of a single BEC fragment, we 
find that the variational analysis is applicable and produces results in 
good agreement with numerical simulation. Our investigation shows that 
the chemical potential (energy) of stationary localized BEC is related 
to the average height of the random potential. 
The larger  the average height 
is, the larger the chemical potential (energy) is. For a weakly disordered 
potential, the localized BECs are found to have an exponential tail as 
expected in weak Anderson localization \cite{NJP-10-045019}. 
The exponential tails are asymmetric 
and are modulated by the disordered potential. The effects of 
nonlinearity on localization are investigated carefully. For a
sufficiently weak nonlinearity, the increase of the repulsive 
nonlinearity leads to an increase of the localization length of the 
exponential tail and the increase of the chemical potential (energy) of 
stationary localized BEC. For larger nonlinearity, the localization will  be destroyed and the 
BEC will be  in a multi-fragmented state. We 
also investigated the expansion of a noninteracting BEC in the 
disordered potential. We find that the BEC will be locked in an 
appropriate localized state after an initial quick expansion and will execute 
breathing oscillation around a mean shape of the localized state when a 
BEC at equilibrium in the harmonic trap is suddenly released into a
potential with weak disorder.

\acknowledgments
%\ack

FAPESP and CNPq (Brazil) provided partial support. 
%Y. S. Cheng
%undertook this work with the support of the post-doctoral  program
%of FAPESP (Brazil) and the Science and Technology Program of the
%Education Department of Hubei, China.

\hskip .2 cm

%\section*{References}

\end{document}